\documentstyle[floats, aps, epsfig]{revtex}

\begin{document}

\twocolumn[\hsize\textwidth\columnwidth\hsize\csname @twocolumnfalse\endcsname

\title{Thermodynamic Compressibility of a Two-Dimensional Electron System:
Signature of a Droplet State}

\author{S. C. Dultz, B. Alavi, and H. W. Jiang }
\address{Department of Physics and Astronomy, University of California at
Los Angeles, Los Angeles, CA 90095}

\date{\today}

\maketitle

\begin{abstract}
We have used a field-penetration method to measure thermodynamic
compressibility of a moderately interacting two-dimensional electron system ($r_{s}$ $\approx$ 0.5-3) in a
three terminal
GaAs/AlGaAs device, fabricated with an epitaxial lift-off technique. We found
that the density and temperature dependencies of the compressibility are qualitatively different from that observed in earlier studies of the 2D hole system, where interaction energies are considerably stronger.  We show that the observed characteristics can be described by the recently developed formalism for compressibility of the droplet state.
\end{abstract}

]

\narrowtext
The apparent metal-insulator transition discovered in low-density, high
mobility Si, SiGe and
GaAs based two-dimensional semiconductor systems continues to capture much interest in the general condensed matter physics community \cite {review}.  Although the experimental work in this research area has been predominately transport studies, there is recent work,
both experimental \cite{Dultz,Ilani1,Ilani2} and theoretical
\cite{Varma,Xie} devoted to the
study of the thermodynamic compressibility of the
interaction of disordered two-dimensional systems, in the context of the 2D
metal - insulator
transition.

In our earlier measurement of a strongly correlated two-dimensional hole
system (2DHS) in a
p-type GaAs/AlGaAs heterostructure, we found that the negative
inverse-compressibility at
low-densities reaches a minimum value at the apparent metal-insulator
transition and than
increases dramatically with further decreasing density.  The results show a
strong
correlation between the qualitative change in thermodynamic compressibility
and the
transport signature of the 2D metal-insulator transition. In that
experiment, the 2DHS has a rather strong correlation due to the heavy
effective mass.
The dimensionless parameter $r_{s}$ which measures the ratio of the
correlation energy
to the Fermi energy ranges between 5 to 25 and the qualitative change in the
compressibility
was observed at $r_{s}$=13.  To understand whether this behavior is specific
to strongly
correlated systems and how it is linked to the 2D metal-insulator
transition, we have studied
another two-dimensional electron system
(2DES) where interactions are much weaker, $r_{s}$ = 0.5-3.  There is no
transition in
the transport. We found that the compressibility behavior in the present
system is
qualitatively different from that observed in the 2DHS system.  Its
density
dependence can be well described by the theory \cite{Xie} of percolating droplet states.

To measure the thermodynamic compressibility, we have used a field
penetration technique,
developed initially by Eisenstein.\cite {Eisenstein} Unique to our setup,
the 2DES is
sandwiched between two metal electrodes (the front and back gate) as shown
in the inset of
Fig. 1. By using an epitaxial lift-off technique for removing the MBE
substrate, we were able
to place a back-gate at a distance of 1000 nm from the 2DES, while the
front gate
is 167 nm away. The samples used in this
experiment were cut from an n-type GaAs/Al$_{x}$Ga$_{1-x}$As wafer
fabricated by molecular
beam epitaxy. A 70 \AA~Al$_{.30}$Ga$_{.70}$As undoped spacer was used with an atomic
planar doping
layer of Si just above with a concentration of $1.5\times 10^{12}\
cm^{-2}$. The rest of
the sample, designed specifically to do the substrate removal, was identical
to the one used
for the 2DHS compressibility study \cite {Dultz}. The mobility of the sample was $186,000\ cm^{2}/V s$ at a density of about $n=6\times
10^{11}\  cm^{-2}$ when the sample was unbiased. The wafer was carefully designed so that the mobility of the carriers was comparable to that
in the 2DHS studied earlier.

\begin{figure}[h]
\begin{center}
\includegraphics[scale=0.35, angle=-90]{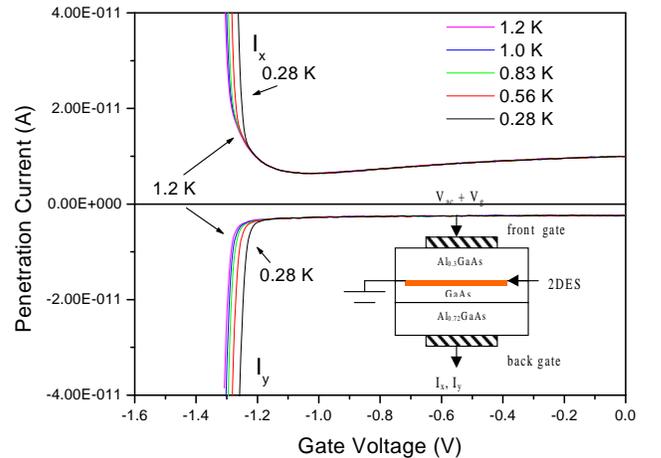}
\caption{Penetration current components $I_{x}$ and $I_{y}$ vs gate voltage
for five temperatures at an excitation frequency of 100 Hz: the temperatures
are labeled from right to left as 0.28 K, 0.56 K, 0.83 K, 1.0 K and 1.2 K
respectively. Inset: the schematic of the experimental setup.}
\end{center}
\end{figure}

To perform the experiment, we applied a 10 mV AC excitation voltage
$V_{ac}$ to the front gate.  A DC voltage $V_{g}$ was superimposed to control
the carrier density.  With the 2DES grounded, the penetration current from the front gate to
the back gate was detected by a lock-in amplifier at an excitation
frequency of 100 Hz.
The frequency was chosen in such a way as to maximize our signal-to-noise
ratio while remaining
in a frequency independent regime.

As described before\cite {Dultz}, by modeling the system as a distributed circuit, both the quantum capacitance $C_{q}$ and the resistance $R_{s}$ of the channel for the 2D carriers could be extracted individually based on the measured values for the in-phase $I_{x}$ and $90^{o}$ phase $I_{y}$ current components:

\begin{equation}
I={i\omega C_{1}C_{2}V_{ac}\over C_{1}+C_{2}}
\bigg[1-\bigg({C_{q}\over C_{1}+C_{2}+C_{q}}
\bigg){\tanh (\alpha )\over \alpha }
\bigg],
\end {equation}
\begin{equation}
\alpha =\sqrt {i\omega {C_{q}(C_{1}+C_{2})\over C_{1}+C_{2}+C_{q}}R_{s}}
.
\end {equation}

where $\omega$ is the frequency of the excitation voltage, and $C_1$ and $C_2$ are the geometric capacitances between the front and back gates with respect to the 2D layer.

The quantum capacitance per unit area $c_{q}$ is related to the compressibility, $\kappa=\big({dn\over d\mu}\big)/n^{2}$, by $c_{q}= e^{2}\big({dn\over d\mu}\big)$, where $\mu$ is the chemical potential and $n$ is the carrier density.

Figure 1 shows both the in-phase $I_{x}$ and $90^{o}$ phase $I_{y}$
components of the penetrating current as a function of the gate voltage for
five different temperatures ranging from 0.28 K to 1.2 K.  The d$\mu$/dn value, which is inversely proportional to the compressibility,
is extracted and plotted in Figure 2 as a function of the $r_{s}$ value in the bottom axis.  At 100 Hz, d$\mu$/dn is almost directly proportional to $I_{x}$. There are important differences between this curve and the same curve for the 2DHS.
First, it is immediately apparent from the data in Figure 1 that the
resistive component
of the signal does not have a temperature independent crossing point like
that for the 2DHS.  This should be expected from a sample with only weak to
strong
localization crossover behavior as shown in the transport measurements.  The
transport
properties of the 2DES can be understood in terms of the more
conventional scaling theory. \cite{Dahm}  Secondly, d$\mu$/dn for the 2DES
gas is positive
everywhere. At high densities, this dependence is expected.  As one can see
from the
quantitative agreement with the Hartree-Folk (HF) approximation calculation
for low
$r_{s}$ values.\cite {Tanatar}.  Since the effective mass of electrons in
GaAs
is only 0.067 times the rest mass of an electron (roughly 5 times less than
the
effective mass of the holes in GaAs), the maximum $r_s$ value of this
particular system is
only about 3 for the lowest density studied here.
Positive compressibility means that the kinetic
energy is always greater than the exchange energy for the present system.
Third, although
there is still a minimum in d$\mu$/dn, just like that observed in
the 2DHS, the turn-around is not nearly as sharp as is the case for that system and occurs
at a density of roughly ($r_{s}=1.4$). Finally, the temperature
dependence of d$\mu/dn$ is rather strong, in contrast to what is observed in
the 2DHS where d$\mu/dn$ is temperature independent for a large range of temperatures from 0.3 K to 4.2 K.

\begin{figure}[h]
\begin{center}
\includegraphics[scale=0.35, angle=-90]{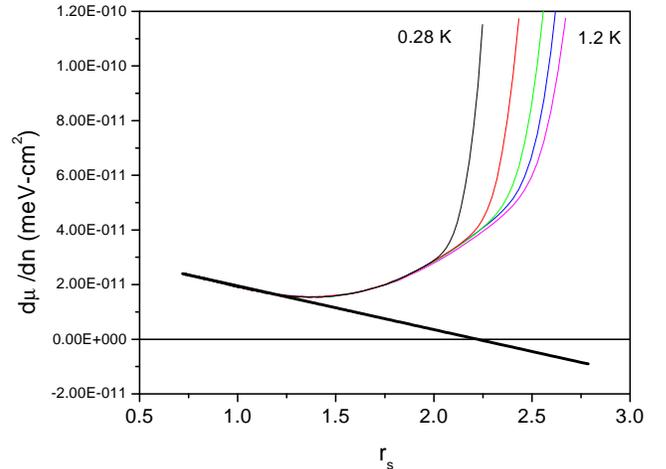}
\caption{$d\mu/dn$ is plotted against $r_{s}$ for the data in Figure 1.  The
solid line is calculated using HF approximation.}
\end{center}
\end{figure}

Over the years, it has been suggested that the interacting 2D electrons in the
presence of
disorder can be understood by the droplet, or puddle, model. \cite{He,Meir}  In this model, the
electrons are
separated into ``liquid'' droplets with local density higher than the average
density and
``gas'' islands with lower density.  For a given disorder, due to the potential
fluctuation of the donors,
as the Fermi energy is decreased, the ``gas''
region increases.  In this model, a cross-over from metallic behavior to insulating
behavior should be seen around the percolation threshold. \cite{Efros}

Specifically, the thermodynamic compressibility has been recently calculated by Shi and Xie for an interacting system within the local density approximation (LDA).\cite{Xie}  The LDA result produced a ``turn-around'' in $d\mu/dn$ which appears to be similar to that observed in the 2DHS system. \cite{Dultz} Furthermore, the LDA calculation was in good agreement with numerical simulations using ``liquid'' droplets and ``gas'' islands.

Following the work of Shi and Xie, we have calculated the compressibility
numerically using LDA for the experimental parameters and found that the basic characteristics of the electron data can be well described by the formalism of the LDA.

In the LDA calculation, the local density ($n_{eff}$) rather than the
average density, is used
to determine the total energy of the system.  The local density is assumed
to be
$n_{eff}=n/f(n)$, where $f(n)$ is the fractional area of the high density region.  The chemical potential is therefore:

\begin{equation}
\mu(n)= d(\varepsilon_{0}(n_{eff})n)/dn,
\end{equation}

where $\varepsilon_{0}$ is the energy density for the uniform electrons and
is given by\cite{Tanatar}:

\begin{equation}
    \varepsilon ={1\over 2}\varepsilon_{Fermi}+\varepsilon _{exchange}={\pi
\hbar ^{2}n\over 2m^{*}}
    -{4\over 3}\sqrt {2\over \pi }{e^{2}\over 4\pi \epsilon }n^{1\over 2},
\end{equation}

where $\epsilon$ is the dielectric constant of the host material and $m^{*}$
is the effective mass of the 2DES.

The chemical potential and $d\mu/dn$ are calculated numerically. As shown in
Figure 3, the
LDA approximation describes the data rather well in the high temperature
limit. To fit the
data we use the following relation for $f(n)$:

\begin{equation}
f(n)= {1\over(1+(n_{0}/n)^{\alpha})}
\end{equation}

$n_{0}$=$1.6\times 10^{11}\  cm^{-2}$ and $\alpha$=2.0 have been used to
produce the
theoretical curve in the figure.  It turns out that the numerical fit is very
sensitive to the
value of $\alpha$ and $n_{0}$.  Small deviations of about 10 percent from
$n_{0}$=$1.6\times 10^{11}\  cm^{-2}$ or $\alpha$=2.0 dramatically effect the curve in such a way that we can no longer qualitatively fit the data.  Incidentally, the value of $\alpha$ of 2.0 is very close to that of 2.3 used to fit the numerical simulation by Shi and Xie \cite{Xie} for off-plane, un-correlated charge impurities and this parameter should depend on the form of the disorder potential.  We do not understand why the data at lower temperatures or lower densities deviate from the theoretical
curve.  It is possible that the temperature dependence of the size of the lower density islands is allowing the field to penetrate to the back gate more easily at the lowest temperatures.  This could create a temperature dependence in $f(n)$ in equation 5, which could explain a temperature dependent $d\mu/dn$ curve.  However, it is unlikely that the temperature dependence would come about from a single temperature dependent parameter in the above equation. \cite {Xie2}

\begin{figure}[h]
\begin{center}
\includegraphics[scale=0.35, angle=-90]{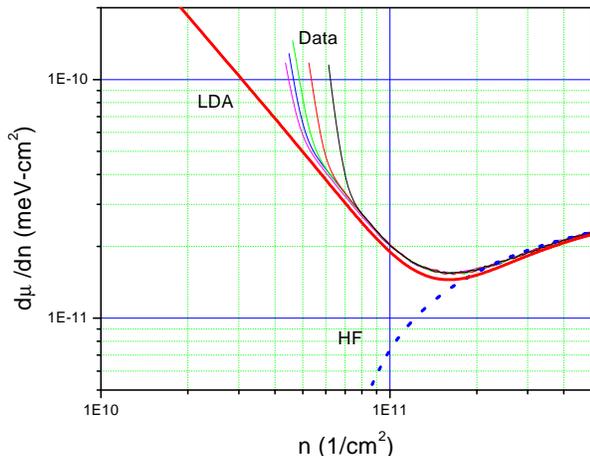}
\caption{The data is compared with the LDA approximation.  The LDA
approximation describes the data well in the high temperature limit.
$n_{0}$=$1.6\times 10^{11}\  cm^{-2}$ and $\alpha$=2.0 are used for the
fitting.}
\end{center}
\end{figure}

It is important to note that the accurate data fit in Figure 3 to
the theoretical
curve should not give the impression that the droplet model explains all experimental data for different
systems.  In fact, Figure 4 shows that the LDA does not fit the data for the
2DHS density
dependence of $d\mu/dp$ at all.  To compare with the data in reference
\cite{Dultz} we cannot find parameters of $\alpha$ and $n_{0}$ which would
fit the curve even qualitatively.  Incidentally, we can match the data
with an expression which contains a mathematical singularity,
which of course is inconsistent with that for percolation.  As an example, we plot a curve of the LDA using $f(n)=(1-(n_{0}/p)^{\beta})$. A reasonable fit is obtained with $n_{0}$=
$3.5\times
10^{10}\  cm^{-2}$ and $\beta$ = 0.06.  This observation implies
that the physics of
the 2DHS which appears in the large $r_{s}$ regime, CANNOT be described by
the droplet model.

  The necessity of a singularity in f further suggests that there
is a phase transition in the 2DHS system.  This
observation further supports the notion that the non-analytical behavior of
the thermodynamic
compressibility is consistent with the unusual signature of the
metal-insulator transition
observed in the transport for the same sample. Indeed there are theories
describing the 2D
metal-insulator transition as a phase transition from a liquid to a Wigner
glass state\cite {Chakravarty,Paster} or from a paramagnetic state to a spontaneously ordered ferromagnetic sate \cite{Chamon}.  It is difficult to determine the existence of these potential states from these measurements due to the absence of a quantitative analysis of the compressibility of the Wigner glass state or the ordered ferromagnetic state.  For example, the compressibility of a Wigner crystal, in the absence of disorder, is known theoretically \cite{Shklovskii} to be only twice that for a
liquid in the HF theory.  However, compressibility for a glass state, in the presence of realistic disorder, has not yet been calculated.

\begin{figure}[h]
\begin{center}
\includegraphics[scale=0.35, angle=-90]{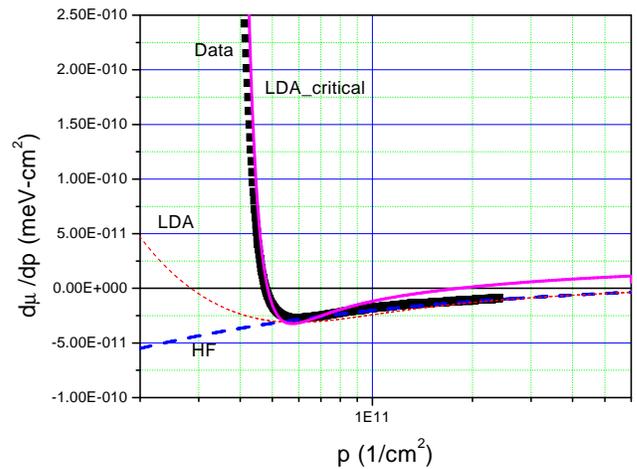}
\caption{In contrast, the LDA approximation cannot describe the $d\mu/dp$
vs. $p$ curve
obtained in the 2D hole system, since the theoretical curve varies slowly
with $p$.
A reasonable fit can only be obtained
if there is a singularity in the filling factor $f(n)$, as described in the
text.}
\end{center}
\end{figure}

It is reasonable to speculate about why the droplet model is more
appropriate to describe the smaller $r_{s}$ system. It is now a rather common view that the interaction can suppress the effect of disorder.  As shown in the numerical simulation, as $r_{s}$ becomes more important the system tends to keep the electron density as uniform as possible to minimize the Coulomb interaction. So for the strongly interacting heavy mass hole system, it is possible that the droplet state is largely suppressed.

In summary, we have measured the compressibility of a moderately
interacting 2D electron system ($r_{s}$ $\approx$ 0.5-3). We revealed
qualitative differences between this system and that observed earlier in a 2D hole system, where interactions are considerably stronger. We found that the compressibility data for the 2DES can be described in the framework of the droplet state model using the local density approximation.

We would like to thank M. Sarachik and X. C. Xie for useful discussions.  This work is supported by NSF under grant \# DMR 0071969.

\end{document}